\documentstyle[prd,aps,preprint]{revtex}
\begin{document}
\draft

%
%

\preprint{Nisho-02/1} \title{PseudoSkyrmion Effects on Tunneling Conductivity
in Coherent Bilayer Quantum Hall States at $\nu =1$}
\author{Aiichi Iwazaki}
\address{Department of Physics, Nishogakusha University, Shonan Ohi Chiba
  277,\ Japan.} \date{March 1, 2002} \maketitle
\begin{abstract}
We present a mechamism why interlayer tunneling conductivity in coherent bilayer quantum
Hall states at $\nu=1$ is anomalously large, but
finite in the recent experiment. According to the 
mechanism, pseudoSkyrmions causes the finite
conductivity, although there exists an 
expectation that dissipationless tunneling current
arises in the state.
PseudoSkyrmions have 
an intrinsic polarization field perpendicular to the layers,
which causes the dissipation. 
Using the mechanism we show that the large peak in the 
conductivity remains for weak parallel magnetic field,
but decay rapidly after its strength is beyond a critical one,
$\sim 0.1$ Tesla.

\end{abstract}
\hspace*{0.3cm}
\pacs{PACS 73.43.-f,73.21.-b,71.10.Pm,12.39.Dc \\ 
Quantum\,Hall\,Effects,\,\, Skyrmion,\,\,
Interlayer\,Phase\,Coherence
\hspace*{1cm}}
It has recently been observed \cite{cohe} that interlayer
tunneling conductivity in
bilayer quantum Hall states at $\nu=1$ shows
anomalous zero bias peak when layer distance $d$ is comparable with
magnetic length $l=\sqrt{1/eB}$. This suggests \cite{wilczek} that
the condensation of exitonic excitation ( a pair of an 
electron in a layer
and a hole in the other layer ) is realized and the 
interlayer phase coherence between the two layers comes out.
Subsequent observations\cite{gold} of Nambu-Goldstone mode and
quantized Hall drag have strongly supported 
this intriguing feature\cite{iwa} in the
bilayer quantum Hall states. 

It is well known that quantum Hall states in a single layer 
can be understood as a condensed state\cite{kil} 
of bosonized electrons\cite{iwa3}
with a single spin degree of freedom.
In the state, however, there is no physically relevant 
phase associated with the bosonized electrons because
there exists a Chern-Simons gauge symmetry which 
can make the phase 
vanish. On the other hand, in the bilayer quantum Hall states there are 
electrons with two degree of freedom, i.e. pseudo-Spin.
We have, in general, two Chern-Simons gauge symmetries\cite{twoChern}
corresponding to bosonization of these two types of electrons.
They can also make the phases of the electrons vanish.
Hence, there are no physically relevant phases. But some of the bilayer
quantum Hall states \cite{iwa} are described only 
by using a single Chern-Simons gauge symmetry.
The gauge symmetry rotates the phase
of each type of bosonised electrons identically. 
Thus, the difference of the two phases
can not be made to vanish so that it is a
physically relevant variable.
This physically relevant phase difference 
comes out as a result of the condensation
of the exitons. In this way the bilayer quantum Hall
states with the interlayer phase coherence are realized \cite{iwa}. 
Especially, 
the states at total filling factor $\nu_1+
\nu_2=1$ 
are under theoretical and experimental investigation 
at present.

It has been recognized\cite{sawada,macdonald} 
that these bilayer quantum Hall states 
possesses various interesting features corresponding to
parameters present in this system, parallel magnetic field $B_{\|}$,
layer distance $d$, imbalance 
$\sigma =(\rho_1-\rho_2)/(\rho_1+\rho_2)$ of electron densities $\rho_i$
between the layers, etc; the index $i$ denotes $i$-th layer.
Among them, the interlayer phase coherence associated with the condensation
of electron-hole pairs is the most intriguing feature of the states
realized by choosing parameters 
$d \simeq l\,(=\sqrt{1/eB})$ at filling factor $2\pi \rho/eB=1$ ( 
$\rho=\rho_1+\rho_2$ ).
The state is stable against
with changing charge imbalance. This feature is 
associated with the indefiniteness of the electron number
difference in the state, which is a result of the exiton condensation. 
Furthermore, the energies of excitations in the state with $\sigma
\simeq 0$ decrease rapidly with 
the parallel magnetic field, while those of excitations in
the state with $\sigma \simeq 1$ i.e. in a single layer, 
increase with $B_{\|}$\cite{iwa2,sk}. The decrease of the excitation
energy is a property of pseudoSkyrmion\cite{iwa2}, which is
topological excitation with pseudo-spin. On the other
hand, the increase of the excitation energy is that of
Skyrmion\cite{sk} with real spin. 

In the observation of the interlayer tunneling conductance
around $\sigma \simeq 0$, the pseudoSkyrmions are 
relevant excitations and may affect seriously on the tunneling 
current. Although we expect the presence of 
dissipationless tunneling current \cite{iwa} in the state with 
the phase coherence, any small disturbance of the phase coherence 
caused by such excitations may
give rise dissipation in the tunneling current.

In this letter we point out that pseudoSkyrmions induce
the dissipation of the energy of the tunneling current $\vec{J}$.
The point is that the pseudoSkyrmion has a polarization field $\vec{E_p}$
associated with it's polarized electric charge distribution
so that the dissipation $\int \vec{E_p}\vec{J}\neq 0$ occurs. The dissipation by the
pseudoSkyrmion only arises in the states with the imbalance
of the charge densities, $\sigma \neq 0 $. In other words,
the dissipation by the pseudoSkyrmions does not
arise by adjusting exactly the parameter
such as $\sigma =0$.

Let us first derive briefly the Josephson equation in the
presence of the condensation\cite{iwa,macdonald}, $<\Psi_1^{\dagger}\Psi_2>\propto 
e^{-i\theta}\neq 0$,
of the exitons. Hereafter, we consider only the quantum Hall
states with the interlayer phase coherence at $\nu=1$. 
The condensation
is a result of intralayer and interlayer Coulomb interactions
among electrons in the highly degenerate lowest Landau level.
It is easy to see that the condensation naturally 
leads to Josephson equations\cite{iwa}  
governing I-V characteristics of dissipationless currents.
We start with Schrodinger equations of electron fields $\Psi_i$, 
$i\partial_t \Psi_1=v_1\Psi_1-\Delta_{sas}\Psi_2$ and 
$i\partial_t \Psi_2=v_2\Psi_2-\Delta_{sas}\Psi_1$,  
where $v_i$ is a chemical potential of electrons in 
the $i$-th layer and $\Delta_{sas}$ is a tunneling strength.
Note that the kinetic term of electrons is quenched by 
strong magnetic field $B$ perpendicular to the layers.
It follows that the Josephson tunneling current is 
given by 

\begin{equation}
\label{t}
e\partial_t<\Psi_1^{\dagger}\Psi_1>=ie\Delta_{sas}(<\Psi_1^{\dagger}\Psi_2>
-<\Psi_1\Psi_2^{\dagger}>)=2e\Delta_{sas}\sqrt{\rho_1 \rho_2}\sin \theta .
\end{equation}
where 
we have used the normalization of the field determined from
more detail calculations\cite{iwa,macdonald}; 
$<\Psi_1^{\dagger}\Psi_2>\simeq \sqrt{\rho_1\rho_2}\,e^{-i\theta}$.
Similarly, we obtain the other equation
governing the development of the phase $\theta$,
$\partial_t \theta=(v_1-v_2)=eV $. Therefore, it is very natural to
expect that Josephson effects arise in this 
coherent bilayer quantum Hall system. In the above derivation,
however, the effect of pseudoSkyrmions is not included.
As far as the filling factor is exactly equal to $1$,
such Skyrmion excitations are absent in the quantum Hall states.
But, in general, the Skyrmions are present in 
the state with $\nu \neq 1$. It seems apparently that their effects
are negligible because their number is relatively so small
that the effects are of the order of $\delta \nu=\nu -1$;
we can choose any small values of $\delta \nu$, e.g. $1/100$.
We should mention, however, that the phase coherence is needed for
dissipationless tunneling current, but the presence of the 
pseudoSkyrmions disturbs the coherence because of 
their distorted phase configuration. Therefore, even if the 
number of the pseudoSkyrmions are negligibly small,
we need to take account of their effects on the conductivity.

As we have shown in the previous paper\cite{iwa2}, solutions of the pseudoSkyrmions in
bosonized electron theory\cite{iwa3} of quantum Hall states are given by

\begin{equation}
\label{sol}
\Psi_1=\sqrt{\rho_1}(z+c_1)\exp(-a(r))\exp(i\theta_1) \quad \mbox{and}
\quad \Psi_2=\sqrt{\rho_2}(z+c_2)\exp(-a(r))\exp(i\theta_2),
\end{equation}
with $z=x-iy$ and $r=|z|$,
where bosonized electron field $\Psi_i$ in the i-th layer
goes to,
$\sqrt{\rho_i}\exp(\theta_i)$, as $r \to \infty$; $\theta=\theta_1-\theta_2$.
The function $a(r)$ is approximately given by
$\exp(-a(r))\simeq \sqrt{1/(r^2+c^2)}$ for $c>l$. Here the parameter $c$ represents
length scale of the pseudoSkyrmion; $c_1=c\sqrt{\rho_2/\rho_1}$ and 
$c_2=-c\sqrt{\rho_1/\rho_2}$. The scale is determined by
minimizing the energy of the pseudoSkyrmions,

\begin{equation} 
\label{energy}
E_{psk}=4\pi\rho_{ss}+\frac{0.46\,e^2}{\epsilon c} +\frac{0.4\,e^2\sigma^2}{\epsilon l}\,\frac{d}{l}\,
\,\frac{c^2}{l^2}  
+\frac{\Delta_{sas}\sqrt{1-\sigma^2}}{2\pi}\,\frac{c^2}{l^2},
\end{equation}
with $\rho_{ss}\simeq 0.005\,(1-\sigma^2)
e^2/\epsilon l$ for $d/l=1.5$ and dielectric constant $\epsilon$. 
The first term represents exchange energy of
pseudo-spin, the second one does Coulomb energy, the third one
does charging energy and the final term represents tunneling energy.
Thus, the typical scale of pseudoSkyrmion $c$ is given by 

\begin{equation}
c\simeq l\, \biggl\{\frac{0.46\,e^2/2 \epsilon l}{\Delta_{sas}\sqrt{1-\sigma^2}+
(0.4e^2\sigma^2/\epsilon l)(d/l)}\biggr\}^{1/3}
\end{equation}
Later, we will determine the scale $c$ by using the data\cite{gold} in the 
observation of the $B_{\|}$ dependence of the tunneling conductivity. Then,
we will find that $c\sim 10\,l$ and $\sigma\sim 0.01$.

The pseudoSkyrmion has positive electric charge $|e|$ and is present
in the region of $\delta \nu <0$, on the other hand,
anti-pseudoSkyrmion has negative charge $-|e|$ and is present in
the region of $\delta \nu >0$. Speculated from the fact that 
it possesses the charging energy, the electric charge of the
pseudoSkyrmion on $1$st layer is different from that on $2$nd layer
when $\rho_1-\rho_2\neq 0$.
Hence, it produces a polarization field, $E_p$,
perpendicular to the layers,

\begin{equation}
E_p(\vec{x})=\frac{1}{d\epsilon}\int dy^2
(\frac{1}{|\vec{x}-\vec{y}|}-\frac{1}{\sqrt{(\vec{x}-\vec{y})^2+d^2}})
(\rho_1(\vec{y})-\rho_2(\vec{y}))
\end{equation}  
where $\rho_1(\vec{y})=e(|\Psi_1|^2-\rho_1)$ and $\rho_2(\vec{y})=
e(|\Psi_2|^2-\rho_2)$ with $\Psi_i$ given in eq(\ref{sol}).

This electric field dissipates the energy of the tunneling current
and gives rise to a finite conductivity. To see it we note that
the interlayer tunneling current $J$ involving the effect of a
pseudoSkyrmion is given by

\begin{eqnarray}
\label{J}
J(\vec{x})&=&e\Delta_{sas}\rho \sqrt{1-\sigma^2}\exp(-2a(r))
\bigl\{(r^2-c^2-2cx\frac{\sigma}{\sqrt{1-\sigma^2}}
)\sin\theta +\nonumber \\
   &+&2c y(\frac{1}{\sqrt{1-\sigma^2}})\cos\theta \bigr\},
\end{eqnarray}
where we have used the formula in eq(\ref{t}) and
the solutions in eq(\ref{sol}).
Then, it follows that the rate $w$ of the energy dissipation is   
   
\begin{equation}
w=\int dx^2 dE_p(\vec{x})J(\vec{x})=2e^2\pi^2\sigma\Delta_{sas}
c^2\rho^{3/2}f(c^2/l^2,d^2/l^2)\sin\theta/\epsilon
\end{equation} 
where we have taken only the term with 
the order of $\sigma$ in the limit as $\sigma \to 0$ and 
$f(c^2/l^2,d^2/l^2)$ is numerically of the order of $0.1$ for $c\simeq 10\,l $, 
$d\simeq 2\times 10^{-6}\mbox{cm}$
and $\rho\simeq 5\times 10^{10}/\mbox{cm}^2$.
We find that the dissipation only arises at $\sigma \neq 0$.
It results from the fact that the pseudoSkyrmion possesses polarization
only when $\sigma \neq 0$. Actually, the charging energy in
eq(\ref{energy}) vanishes at $\sigma =0$.

We now calculate interlayer tunneling conductivity in
the presence of $N_0$ pseudoSkyrmions.  We suppose that
the contribution of each pseudoSkyrmion is incoherently taken 
into account. Then,
The total rate $W$ of the
energy dissipation is $N_0 w$, which is equal to
the product, $W=VI$, of the everage voltage $V$ and 
the total tunneling current $I=I_0+I_s=\int dx^2
(J_0+N_0J_s(\vec{x}))$, where the first term $I_0$ represents
the contribution of the groundstate and
the second term $I_s$ represents
the contribution of the pseudoSkyrmions;
$J_0=2e\sqrt{\rho_1 \rho_2}\Delta_{sas}\sin\theta$,
$J_s(\vec{x})=J(\vec{x})-J_0$ and $I_s=-2\pi N_0e
\Delta_{sas}\rho c^2\sqrt{1-\sigma^2}k_0\sin\theta$
( $k_0$ is a numerical constant ).
 Then, the conductivity $G=\partial
I/\partial V$ is approximately given by

\begin{equation}
G\simeq \frac{I}{V}=\frac{(I_0+I_s)^2}{N_0w}=
\frac{\Delta_{sas}N(1-2\pi \rho c^2
  k_0\delta\nu)^2\epsilon \sin\theta}{2\pi^2c^2\sigma\rho^{3/2}f\delta\nu}\sim
6\times 10^{-5}\Omega^{-1}\frac{(1-2\pi \rho c^2
  k_0\delta\nu)^2\sin\theta}{f\sigma \delta \nu}
\end{equation} 
in the limit of $\sigma \to 0$,
where we have used the parameters taken in the experiment\cite{cohe},
$d/l\simeq 1.6$, $\Delta_{sas}\simeq 0.1\mbox{mK}$ and  
surface area ($\simeq (0.2\mbox{mm})^2$ ) 
occupied by two dimensional electrons. 
$N=\int dx^2\rho$ denotes total number of
electrons; $\delta\nu=N_0/N$
and $k_0$ is of the order of 
$1$.


Although the absolute value of $G$ is much larger than that of the 
observation, the dependence on
$\delta \nu $ agrees roughly with the observation. Even if we take $\delta \nu $
to be $0$, there would be residual pseudoSkyrmions. It seems that 
such residual excitations induce finite conductivity 
observed around $\nu=1$ \cite{cohe}, although there is a possibility
that other mechanisms of
the current dissipation still work. 
We note that the dependence of $G$ on the imbalance parameter 
$\sigma $ is the same as that on $\delta \nu$. It may be easy to check
the dependence observationally. 
We also note that in 
our formula both of $G$ and $I$ are proportional to $\sin \theta$.
Therefore, it is interesting to check this point by 
an experiment with current feed.

Now, let us consider the dependence of the conductivity on parallel
magnetic field $B_{\|}$ pointing to $y$ direction. The effect is taken
only by changing
the phase factor $\theta $ in the currents $J$ in eq(\ref{J})  
and $J_0$ such that
$\theta \to \theta + Qx$ with $Q=edB_{\|}$; 
the polarization field $E_p$
is not changed.
Then, we see that 
the current $I_s(Q)=\int dx^2 J_s(\vec{x}, Q)$ involving the 
effect of pseudoSkyrmion decays such as 
$I_s(Q) \to \exp(-Q c)$ for $Q \to \infty$. We also note that 
the groundstate current $I_0(Q)=\int dx^2 J_0(Q)$ behaves such as
$I_0(Q) \propto \sin(e\Phi/\Phi_0)/(e\Phi/\Phi_0)$ where
$\Phi =d\int dx  B_{\|}$ denotes flux penetrating the region 
between the two layers; $\Phi_0=e/\pi$. This term is negligibly
small even for $B_{\|}=0.1$T, while the other term $I_s(Q)$
is not small for such $B_{\|}$. Therefore, the conductivity behaves
such that 
$G(Q) \sim (I_0(Q)+I_s(Q))^2/(N_0w(Q)) \sim I_s(Q)^2/(N_0w(Q)) \to \exp(-Qc)$ as $Q$ becomes
large. Comparing this one with the observation, we can determine the
scale of pseudoSkyrmion, $c\sim 10\,l$.
This value agrees roughly with the one derived with 
a different method \cite{moon}. 
We find that the pseudoSkyrmion in the observation is 
large so that even if we take $\delta\nu = 1/100$, the whole plain is 
occupied by the pseudoSkyrmions.

We stress that not only the presence of the zero bias peak even at
small $B_{\|}\neq 0$,
but also the rapid decay of the peak as $B_{\|}$ going beyond $\sim
0.1$ Tesla, 
can be explained as the effect of the pseudoSkyrmions. 
This fact strongly suggests that the finite interlayer tunneling conductivity 
in the recent observation is 
caused mainly by the pseudoSkyrmions,
although our estimation of the conductivity is much larger than that of 
the observation.

Finally we wish to point out that
in order to check our conclusion and observe dissipationless Josephson 
tunneling current, 
the imbalance $\sigma =(\rho_1-\rho_2)/\rho$ 
should be diminished
with carefully adjusting bias voltage between the two layers.
Then, we expect that the balance state $\sigma=0$ would be
realized automatically when $\sigma $ is adjusted to be less than a 
critical one, and   
we will be able to see Josephson effects in 
the quantum Hall states even in the 
presence of residual pseudoSkyrmions.

We thank F.Z. Ezawa and A. Sawada for fruitful discussions 
and members of theory group in KEK for their
hospitality.





\end{document}